\documentclass[aps,prb,twocolumn,superscriptaddress,floatfix]{revtex4-1}

\usepackage{graphicx,graphics}
\usepackage{dcolumn}
\usepackage{amsmath,amssymb,amsfonts}
\usepackage{latexsym,verbatim}
\usepackage{bm}
\usepackage{color}
\usepackage{ulem}
\usepackage[percent]{overpic}
\usepackage[breaklinks=true,colorlinks,citecolor=blue,linkcolor=blue,urlcolor=blue]{hyperref}

\begin{document}
\title{Nonlocal topological valley transport at large valley Hall angles}
\author{Michael Beconcini}
\email{michael.beconcini@sns.it}
\affiliation{NEST, Scuola Normale Superiore, I-56126 Pisa,~Italy}
\author{Fabio Taddei}
\affiliation{NEST, Istituto Nanoscienze-CNR and Scuola Normale Superiore, I-56126 Pisa,~Italy}
\author{Marco Polini}
\affiliation{Istituto Italiano di Tecnologia, Graphene Labs, Via Morego 30, I-16163 Genova,~Italy}
\begin{abstract}
Berry curvature hot spots in two-dimensional materials with broken inversion symmetry are responsible for the existence of transverse valley currents, which give rise to giant nonlocal dc voltages. Recent experiments in high-quality gapped graphene have highlighted a saturation of the nonlocal resistance as a function of the longitudinal charge resistivity $\rho_{{\rm c}, xx}$, when the system is driven deep into the insulating phase. The origin of this saturation is, to date, unclear. In this work we show that this behavior is fully compatible with bulk topological transport in the regime of {\it large} valley Hall angles (VHAs). We demonstrate that, for a fixed value of the valley diffusion length, the dependence of the nonlocal resistance on $\rho_{{\rm c}, xx}$ weakens for increasing VHAs, transitioning from the standard $\rho^3_{{\rm c}, xx}$ power-law to a result that is independent of $\rho_{{\rm c}, xx}$. 
\end{abstract}

\maketitle

{\it Introduction.---}Two-dimensional (2D) materials with no inversion symmetry like gapped graphene~\cite{gorbachev_science_2014,sui_naturephys_2015,shimazaki_naturephys_2015} and transition metal dichalcogenides~\cite{dixiao_prl_2012,mak_science_2014,lee_naturenano_2016} display the so-called valley Hall effect (VHE)~\cite{dixiao_prl_2007}, i.e. charge-neutral valley currents that flow transversally to an applied electric field. Such intrinsic VHE is enabled by the existence of regions in the Brillouin zone where the band structure of a material hosts Berry curvature~\cite{dixiao_prl_2007,dixiao_rmp_2010,nagaosa_rmp_2010,lensky_prl_2015} hot spots with opposite signs. A similar effect in the spin channel, the spin Hall effect, has attracted a great deal of interest and has been extensively investigated in semiconductors and metals in the last decade~\cite{sinova_rmp_2015}.

\begin{figure}[t]
\centering
\includegraphics[width=1.0\columnwidth]{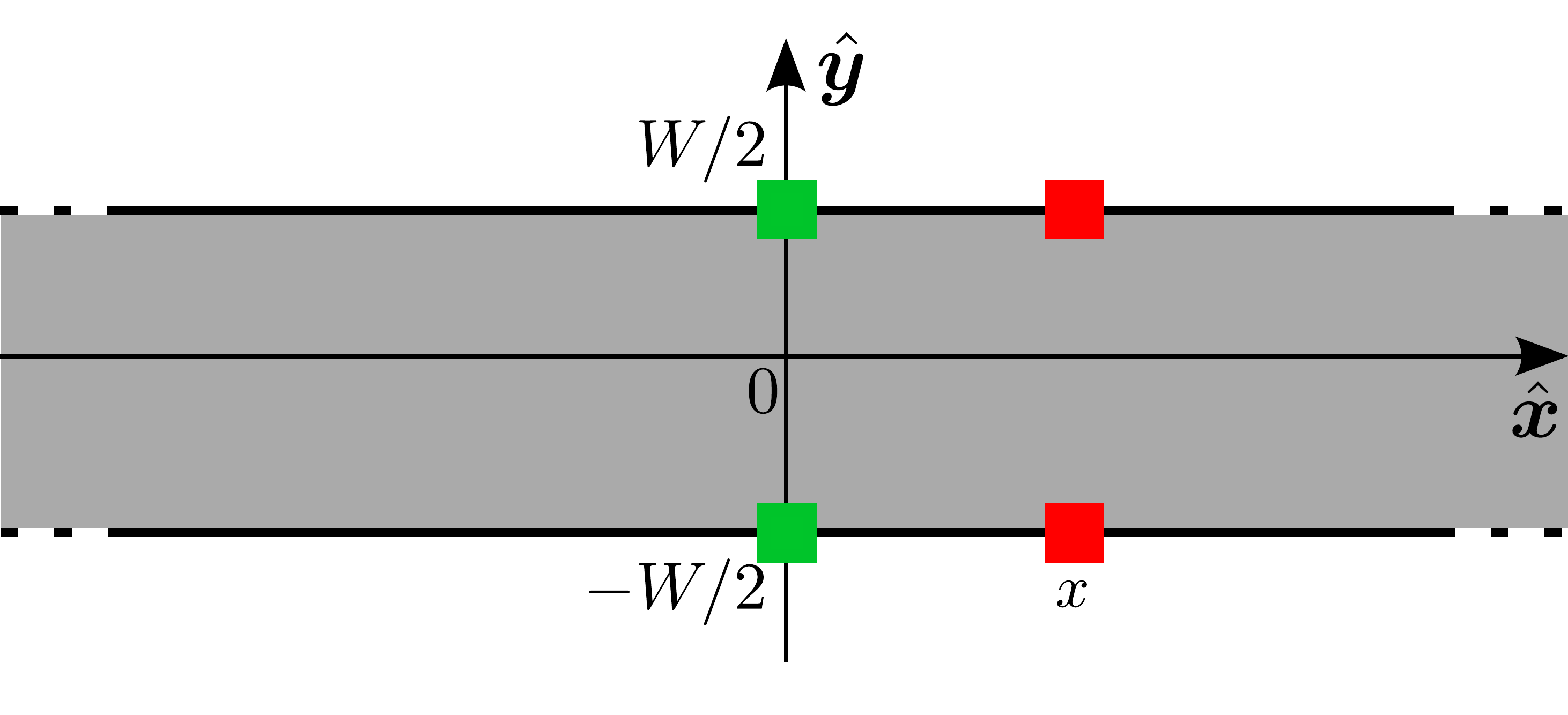}
\caption{(Color online) A sketch of the nonlocal transport setup analyzed in this work. A charge current $I$ is driven between injector and collector electrodes (green) that are in electrical contact with a 2D conductive channel (grey-shaded area). A steady-state dc voltage $\Delta V(x)$ is established between two probe electrodes (red) at a lateral distance $x$ due to diffusion of valley currents (i.e.~valley Hall effect) and their subsequent conversion into a regular charge voltage (inverse valley Hall effect). The nonlocal resistance is defined by $R_{\rm NL}(x) = [\phi(x,-W/2)-\phi(x,+W/2)]/I\equiv \Delta V(x)/I$, where $\phi(x,y)$ is the 2D electrical potential.\label{fig:setup}}
\end{figure}

Nonlocal transport~\cite{valenzuela_nature_2006,abanin_prb_2009,brune_naturephys_2010,abanin_science_2011,balakrishnan_naturephys_2013,wang_prl_2015} is a particularly useful tool to probe the existence of long-range spin and valley Hall transport. A typical nonlocal experimental setup is depicted in Fig.~\ref{fig:setup}. In this geometry, one applies a current bias $I$ between two electrodes on the opposite sides of a 2D conductive channel and measures a steady-state voltage drop $\Delta V(x)$ at a lateral distance $x$ from the current injection path. The ratio $R_{\rm NL}(x) \equiv \Delta V(x)/I$ defines a nonlocal resistance, which depends on the valley Hall angle (VHA) $0\leq \theta_{\rm VH} < \pi/2$ defined by
\begin{equation}\label{eq:VHA}
\tan(\theta_{\rm VH}) \equiv \frac{\sigma_{{\rm v}, xy}}{\sigma_{{\rm c}, xx}}~.
\end{equation}
Here, $\sigma_{{\rm v}, xy}$ is the transverse valley Hall conductivity, $\sigma_{{\rm c}, xx}$ is the ordinary longitudinal charge conductivity. For a fixed value of the valley diffusion length $\ell_{\rm v}$---see below Eq.~(\ref{eq:valley_diffusion_length})---and in the limit $\theta_{\rm VH}\ll 1$, it has been shown~\cite{gorbachev_science_2014,abanin_prb_2009} that $R_{\rm NL}(x)\propto \rho^3_{{\rm c}, xx}$, where $\rho_{{\rm c}, xx}=1/\sigma_{{\rm c}, xx}$ is the longitudinal charge resistivity.

Recently, topological valley currents have been detected in high-quality gapped graphene~\cite{gorbachev_science_2014,sui_naturephys_2015,shimazaki_naturephys_2015}. Early experimental work~\cite{gorbachev_science_2014} carried out in aligned stacks of single-layer graphene and hexagonal boron nitride reported clear evidence of order-one VHAs. Subsequent nonlocal transport measurements in encapsulated bilayer graphene in the presence of a perpendicular displacement field $D$ reported~\cite{sui_naturephys_2015,shimazaki_naturephys_2015} clear departures from the $R_{\rm NL}(x)\propto \rho^3_{{\rm c}, xx}$ law. In particular, large deviations from the cubic power-law were experimentally detected in large displacement fields and/or at large temperatures. While departures from the cubic law at large temperatures may be, at least partly, linked to inelastic effects~\cite{beconcini_arxiv_soon}, deviations at intermediate temperatures but large displacement fields have a greater chance of originating from large values of the VHA. For example, Shimazaki et al.~\cite{shimazaki_naturephys_2015} observed a clear saturation of $R_{\rm NL}(x)$ as a function of $\rho_{{\rm c}, xx}$ for increasing $D$ and at a fixed temperature ($T=50~{\rm K}$)---see Fig.~4 in Ref.~\onlinecite{shimazaki_naturephys_2015}. Motivated by this body of experimental literature, we here present a theoretical study of non-local coupled charge-valley transport at arbitrary values of the VHA. We find that the aforementioned saturation is fully compatible with bulk topological valley transport occurring at {\it large} VHAs.
{\it Theory of nonlocal transport at arbitrary VHAs.---}We consider a generic 2D material that supports valley-polarized transport in response to an external charge current $I$. We introduce the equilibrium number densities $n_{\xi}$  where $\xi=K$, $K^\prime$  is the valley index. Below, we only present analytical results for the situation in which $n_{K} = n_{K^\prime}$ at equilibrium (i.e.~for $I=0$). Generalized expressions for the case $n_{K} \neq n_{K^\prime}$ can be obtained in a straightforward manner.

Because of the applied current and Berry curvature hot spots~\cite{dixiao_prl_2007,lensky_prl_2015} with opposite signs in the two valleys $K, K^\prime$, an excess valley polarization $\delta n_{\rm v}({\bm r})$ is established to first order in $I$ and under steady-state conditions. Here, $\delta n_{\xi}({\bm r})=n_{\xi}({\bm r}) - n_{\xi}$ 
and $\delta n_{\rm v}({\bm r}) = \delta n_{K}({\bm r}) - \delta n_{K^\prime}({\bm r})$. 
Similarly, we denote by $\delta n({\bm r}) = \delta n_{K}({\bm r}) + \delta n_{K^\prime}({\bm r})$ the excess carrier density. We now lay down a theory that allows us to calculate $\delta n_{\rm v}({\bm r})$ and the corresponding electrical potential $\phi({\bm r})$  that is generated due to the inverse VHE {\it away} from the spatial region where $I$ flows.

A theory of nonlocal valley transport starts from a linear-response relation, Eq.~(\ref{eq:LRT}) below, between the current carried by carriers with valley flavor $\xi$ and the gradient of the local electro-chemical potential
\begin{equation}
\Psi_{\xi}({\bm r}) = \phi({\bm r}) -\frac{1}{e}\mu_{\xi}(n_{K}({\bm r}), n_{K^\prime}({\bm r}), T)~.
\end{equation}
Here, $\phi({\bm r})$ is the electric potential in the 2D conductive channel where electrons roam and $\mu_{\xi}(n_{K}, n_{K^\prime}, T)=\partial f(n_{K}, n_{K^\prime}, T)/\partial n_{\xi}$ where $f(n_{K}, n_{K^\prime}, T)$ is the free energy per unit volume of the homogeneous 2D interacting electron system at temperature $T$. Introducing the homogeneous conductivity matrix $\sigma_{\xi\xi^\prime, ij}$ with Cartesian indices $i,j=x,y$, we immediately find the following relation for the $i$-th spatial component of the current:
\begin{equation}\label{eq:LRT}
-e J_{\xi, i}({\bm r}) = \sum_{j, \xi^\prime}[-\sigma_{\xi\xi^\prime, ij}\partial_{i} \phi({\bm r}) + e D_{\xi\xi^\prime, ij}\partial_{j}n_{\xi^\prime}({\bm r})]~.
\end{equation}
The first term on the right-hand side of Eq.~(\ref{eq:LRT}) represents the drift current in response to the electric field while the second term is the diffusion current associated with the gradient of the valley-resolved densities. We note that the diffusion matrix  satisfies the generalized Einstein relation~\cite{damico_prb_2002}, $e^2 D_{\xi\xi^\prime, ij} = \sum_{\zeta} \sigma_{\xi\zeta,ij}S_{\zeta\xi^\prime}$ with $S_{\xi\xi^\prime}=\partial \mu_{\xi}/\partial n_{\xi^\prime}$  the static valley-stiffness matrix.

For the sake of simplicity, in this work we neglect inter-valley electron-electron interactions (which will be the subject of a separate publication~\cite{beconcini_arxiv_soon}). In this approximation, the {\it off-diagonal} (i.e.~{\it inter-valley}) elements of the conductivity and valley-stiffness matrices {\it vanish} exactly. 

Instead of dealing with the valley-resolved particle current densities, $J_{\xi, i}({\bm r})$ for $\xi= K$, $K^\prime$, 
it is convenient to introduce their sum $J_{{\rm c}, i}({\bm r}) \equiv J_{K, i}({\bm r}) + J_{K^\prime, i}({\bm r})$ and difference  $J_{{\rm v}, i}({\bm r}) \equiv J_{K, i}({\bm r}) - J_{K^\prime, i}({\bm r})$. We find
\begin{equation}\label{eq:charge_current}
-eJ_{{\rm c}, i}({\bm r}) = \sum_{j} [\sigma_{{\rm c}, xx} \delta_{ij} E_j({\bm r})  +e D_{{\rm cv}, xy}\epsilon_{ij}\partial_j\delta n_{\rm v}({\bm r})]
\end{equation}
and
\begin{equation}\label{eq:valley_current}
-eJ_{{\rm v}, i}({\bm r}) = \sum_{j}[\sigma_{{\rm v}, xy} \epsilon_{ij}E_j({\bm r})  +e D_{{\rm v}, xx}\delta_{ij}\partial_j\delta n_{\rm v}({\bm r})]~,
\end{equation}
where $\delta_{ij}$ is the Kronecker delta, 
$\epsilon_{ij}$  is the fully anti-symmetric 2D tensor (i.e.~$\epsilon_{ii}=0$ and $\epsilon_{xy}= - \epsilon_{yx}=1$), $\sigma_{{\rm c}, xx} = 2 \sigma_{KK, xx} = 2 \sigma_{K^\prime K^\prime,xx}$ is the charge conductivity, 
$\sigma_{{\rm v}, xy} = 2 \sigma_{KK,xy} = -2 \sigma_{K^\prime K^\prime,xy}$ is the valley Hall conductivity, $D_{{\rm cv}, xy} = S_{KK}\sigma_{{\rm v}, xy}/(2e^2) =S_{K^\prime K^\prime}\sigma_{{\rm v}, xy}/(2e^2)$, and $D_{{\rm v}, xx} = S_{KK}\sigma_{KK, xx}/e^2=S_{K^\prime K^\prime}\sigma_{K^\prime K^\prime, xx}/e^2$ is the longitudinal valley diffusion constant. 

The first term on the right-hand side of Eq.~(\ref{eq:valley_current}) is responsible for the VHE: an electric field ${\bm E}$ induces a transverse valley current, which results into the accumulation of carriers with opposite valley flavors at opposite edges of the sample. The second term in Eq.~(\ref{eq:charge_current}) in the expression of the charge current describes the {\it inverse} VHE: a valley imbalance $\delta n_{\rm v}({\bm r})$ is converted at the sample edge into an electrical signal, which can be measured away from the current injection path.

The nonlocal resistance stemming from the topological VHE can be calculated by solving the following steady-state equations~\cite{gorbachev_science_2014,abanin_science_2011,abanin_prb_2009}: the continuity equation for the charge current density
\begin{equation}\label{eq:continuity}
\nabla \cdot {\bm J}_{\rm c}({\bm r}) = 0~,
\end{equation}
and the damped diffusion equation for the valley imbalance $\delta n_{\rm v}({\bm r})$
\begin{equation}\label{eq:diffusion}
D_{{\rm v}, xx}\nabla^2\delta n_{\rm v}({\bm r}) -\frac{1}{\tau_{\rm v}}\delta n_{\rm v}({\bm r}) = -\frac{1}{e}\nabla \times [\sigma_{{\rm v}, xy}{\bm E}({\bm r})]~.
\end{equation}
Here, $\tau_{\rm v}$ is a phenomenological relaxation time due to extrinsic inter-valley scattering~\cite{gorbachev_science_2014} and ${\bm E}({\bm r}) = - \nabla\phi({\bm r})$ is the electric field. In deriving the previous equation we have used the local charge neutrality constraint, $\delta n({\bm r})\approx 0$. 

Since $\nabla \times {\bm E}({\bm r})=0$, the term on the right-hand side of the damped diffusion equation (\ref{eq:diffusion}) takes a non-zero value only at the boundary between topologically non-trivial (bulk of the conductive channel) and trivial (outside the conductive channel) regions, where the valley Hall conductivity $\sigma_{{\rm v}, xy}$ changes abruptly in space from a finite value to zero. 

The two differential equations above need to be supplemented by suitable boundary conditions (BCs). The first BC is on the charge current: $-eJ_{{\rm c}, y}(x,y=\pm W/2)= I\delta(x)$, where $W$ is the width of the 2D conductive channel. 
At $x=0$ this BC represents current {\it injection} ({\it extraction}) along the $\hat{\bm y}$ direction---see Fig.~\ref{fig:setup}. For $x\neq 0$, it implies that no charge current can enter or exit the 2D conductive channel away from the injector and collector electrodes. The second BC is $J_{{\rm v}, y}(x,y=\pm W/2)=0$, i.e.~no valley current can flow outside the 2D conductive channel. 

We now note that the divergence of the second term in the right-hand side of Eq.~(\ref{eq:charge_current}) is zero. 
We therefore conclude that the 2D potential $\phi({\bm r})$ obeys the Laplace equation inside the 2D conductive channel:
\begin{equation}\label{eq:Laplace}
\nabla^2\phi({\bm r})=0~.
\end{equation}
However, the BC on the charge current reads as following:
\begin{equation}\label{eq:BC_charge}
[-\sigma_{{\rm c}, xx} \partial_{y}\phi({\bm r}) - e D_{{\rm cv}, xy} \partial_{x} \delta n_{\rm v}({\bm r})]_{y=\pm W/2} = I\delta(x)~.
\end{equation}
{\it This BC therefore formally couples the Laplace equation for the 2D electrical potential $\phi({\bm r})$ with the damped-diffusion equation for $\delta n_{\rm v}({\bm r})$}. The coupling is controlled by the VHA in Eq.~(\ref{eq:VHA}).

The BC on the valley current can be re-written as
\begin{equation}\label{eq:BC_valley}
[\partial_{y} \delta n_{\rm v}({\bm r})]_{y=\pm W/2} = - \frac{\sigma_{{\rm v}, xy}}{e D_{{\rm v}, xx}}[\partial_{x}\phi({\bm r})]_{y=\pm W/2}~.
\end{equation}
\begin{figure}[t]
\centering
\begin{overpic}[width=1\linewidth]{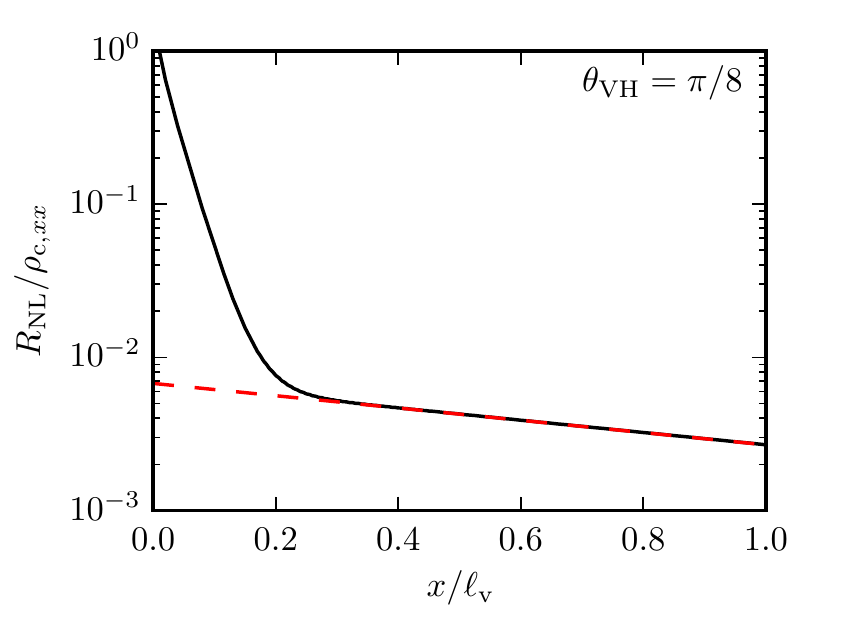}\put(0,70){(a)}\end{overpic}
\begin{overpic}[width=1\linewidth]{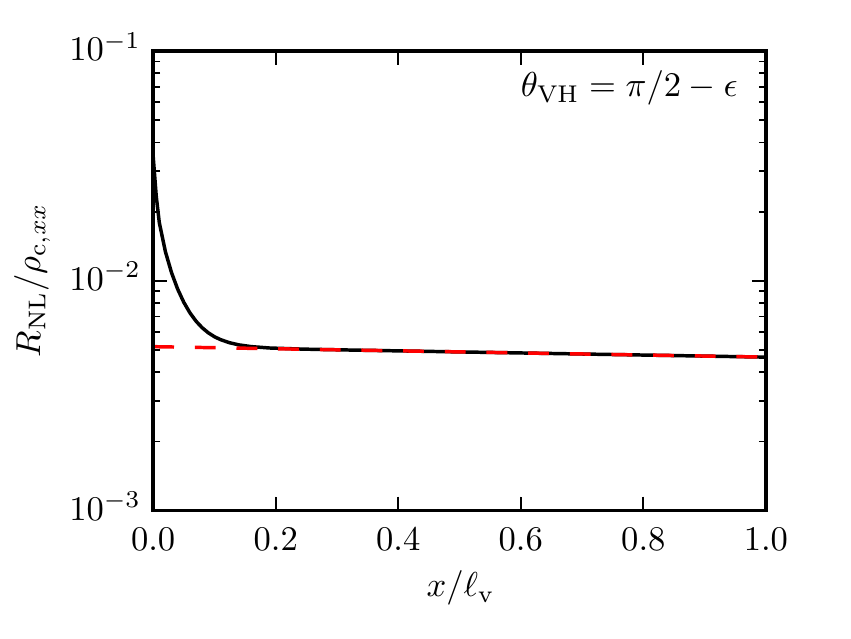}\put(0,70){(b)}\end{overpic}
\caption{(Color online) Spatial dependence of the nonlocal resistance $R_{\rm NL}(x)$ (in units of $\rho_{{\rm c}, xx}$ and in logarithmic scale) for $W/\ell_{\rm v}=0.1$. The analytical result (dashed line) reported in Eqs.~(\ref{eq:non-local-valley-signal-small-Wtilde})-(\ref{eq:effective-diffusion-length}) is compared with the full numerical result (solid line) obtained from Eq.~(\ref{eq:nonlocal-resistance}). Panel (a): $\theta_{\rm VH}= \pi/8$. Panel (b): $\theta_{\rm VH}=\pi/2 - \epsilon$, with $\epsilon= \pi/30$. In panel (a) we clearly notice a bi-exponential behavior, where the first fast exponential decay occurs on the electrostatic length scale $W/\pi$, while the second slow decay occurs on the scale of the valley diffusion length $\ell_{\rm v}$.\label{fig:two}}
\end{figure}
{\it Solution of the problem by Fourier transforms.---}The problem posed by Eqs.~(\ref{eq:diffusion}), (\ref{eq:Laplace}), (\ref{eq:BC_charge}), and~(\ref{eq:BC_valley}) can be solved by Fourier transforming all the unknowns in the longitudinal $\hat{\bm x}$ direction~\cite{torre_prb_2015}:
\begin{equation}\label{eq:FT}
\tilde{f} (k, y) = \int_{-\infty}^{\infty} d x~ e^{-i k x} f(x, y)
\end{equation}
and
\begin{equation}\label{eq:IFT}
f (x, y) = \int_{-\infty}^{\infty} \frac{d k}{2 \pi}~ e^{i k x} \tilde{f} (k, y)~.
\end{equation}
With the aid of Eq.~(\ref{eq:FT}), we can rewrite Eqs.~(\ref{eq:diffusion}), (\ref{eq:Laplace}), (\ref{eq:BC_charge}), and~(\ref{eq:BC_valley}) as following
\begin{equation}\label{eq:diffusion_FT}
[\partial_y^2 - \omega^2(k)] \delta \tilde{n}_{\rm v}(k, y) = 0~,
\end{equation}
\begin{equation}\label{eq:Laplace_FT}
(\partial_y^2 - k^2 ) \tilde{\phi} (k, y) = 0~,
\end{equation}
\begin{equation}
[\sigma_{{\rm c}, xx} \partial_{y}\tilde{\phi}(k , y) +i k e D_{{\rm cv}, xy} \delta \tilde{n}_{\rm v}(k, y)]_{y=\pm W/2} = -I~,
\end{equation}
and
\begin{equation}
[i k \sigma_{{\rm v}, xy}\tilde{\phi}(k , y) + e D_{{\rm v}, xx} \partial_y \delta \tilde{n}_{\rm v}(k, y)]_{y=\pm W/2} = 0~.
\end{equation}
In Eq.~(\ref{eq:diffusion_FT}) we have introduced the quantity $\omega(k) = \sqrt{k^2 + \ell^{-2}_{\rm v}}$, where
\begin{equation}\label{eq:valley_diffusion_length}
\ell_{\rm v} = \sqrt{D_{{\rm v}, xx} \tau_{\rm v}}
\end{equation}
is the valley diffusion length~\cite{gorbachev_science_2014}. Note that the damped-diffusion equation (\ref{eq:diffusion_FT}) and the Laplace equation (\ref{eq:Laplace_FT}) are only coupled by the BCs. 

Seeking solutions of the form $\tilde{\phi}(k, y) = A \cosh (k y) + B \sinh (k y)$ and $\delta \tilde{n}_{\rm v} (k, y) =  C \cosh [\omega(k) y] + D \sinh[\omega(k) y]$, we can determine the coefficients $A$, $B$, $C$, and $D$ by imposing the four BCs at the edges $y=\pm W/2$ of the 2D conductive channel. The final solution is expressed in terms of inverse Fourier transforms. For the steady-state valley imbalance we find
\begin{widetext}
\begin{equation}\label{eq:valley_imbalance}
\delta n_{\rm v} (x, y) = i \frac{I \tan (\theta_{\rm VH})}{e D_{{\rm v}, xx}} \int_{-\infty}^{\infty} \frac{d k}{2 \pi}~e^{i k x} \frac{\cosh [\omega(k)y] }{\omega(k) \coth (kW/2) \sinh [\omega (k) W/2] + k \tan^2 (\theta_{\rm VH}) \cosh [\omega (k) W/2]}~.
\end{equation}
\end{widetext}
The spatial distribution (\ref{eq:valley_imbalance}) of valley polarization may be probed optically. Here, we instead focus our attention on the nonlocal resistance, which we calculate by its fundamental definition: $R_{\rm NL}(x) = [\phi(x,y=-W/2) - \phi(x,y=+W/2)]/I$. We find:
\begin{widetext}
\begin{equation}\label{eq:nonlocal-resistance}
R_{\rm NL}(x) = 2\rho_{{\rm c}, xx}\int_{-\infty}^{+\infty} \frac{dk}{2\pi}~
e^{i k x} 
\frac{\omega(k)/k}{\omega(k)\coth(k W/2) + k \tan^2(\theta_{\rm VH})
\coth[\omega(k) W/2]}~.
\end{equation}
\end{widetext}
Eqs.~(\ref{eq:valley_imbalance})-(\ref{eq:nonlocal-resistance}) are the most important results of this work and are valid for arbitrary values of $\theta_{\rm VH}$. 

For $\theta_{\rm VH} = 0$, we have a purely Ohmic nonlocal signal, which decays exponentially on the length scale $W/\pi$. Straightforward algebra yields
\begin{equation}\label{eq:pure-electrostatics}
R^{(0)}_{\rm NL}(x) = \frac{2\rho_{{\rm c}, xx}}{\pi} \ln\left| \coth \left( \frac{\pi x}{2 W} \right) \right|~,
\end{equation}
which reduces to the well-known van der Pauw formula, $R_{\rm vdP}(x) =4\rho_{{\rm c}, xx} e^{-\pi x/W}/\pi$ for $x \gg W/\pi$.

Illustrative numerical results for $R_{\rm NL}(x)$ obtained from a brute-force fast Fourier transform integration of Eq.~(\ref{eq:nonlocal-resistance}) are reported in Fig.~\ref{fig:two}.

For practical purposes, it is highly convenient to derive an analytical expression for the nonlocal resistance $R_{\rm NL}(x)$ in the limit $x \gg \ell_{\rm v} \gg W/\pi$. This is clearly the regime of interest, where long-range neutral valley currents diffuse over distances that are much larger than the van der Pauw scale $W/\pi$. In this limit, it is possible to obtain a closed-form expression for the nonlocal resistance, which is valid for arbitrary VHAs. Expanding the integrand in Eq.~(\ref{eq:nonlocal-resistance}) for $kW\ll k\ell_{\rm v}$ and integrating we find the following formula
\begin{eqnarray}\label{eq:non-local-valley-signal-small-Wtilde}
\Delta R_{\rm NL}(x) &\equiv& R_{\rm NL}(x) - R^{(0)}_{\rm NL}(x) \nonumber\\
&=&\frac{W}{2 L_{\rm v}}\rho_{{\rm c}, xx}\frac{\tan^2(\theta_{\rm VH})}{1+\tan^2(\theta_{\rm VH})}e^{-|x|/L_{\rm v}}
\end{eqnarray}
for $x \gg \ell_{\rm v} \gg W/\pi$. In Eq.~(\ref{eq:non-local-valley-signal-small-Wtilde}) we have introduced a {\it renormalized} valley diffusion lenght:
\begin{equation}\label{eq:effective-diffusion-length}
L_{\rm v} \equiv \ell_{\rm v}\sqrt{1+\tan^2(\theta_{\rm VH})}~.
\end{equation}
Eqs.~(\ref{eq:non-local-valley-signal-small-Wtilde})-(\ref{eq:effective-diffusion-length}) are compared against the full result (\ref{eq:nonlocal-resistance}) in Fig.~\ref{fig:two} and can be used for analyzing nonlocal transport data at arbitrary values of the VHA.

In the limit of small VHAs, i.e.~for $\sigma_{{\rm v}, xy} \ll \sigma_{{\rm c}, xx}$, Eq.~(\ref{eq:non-local-valley-signal-small-Wtilde}) reduces to
\begin{equation}\label{eq:small_theta}
\lim_{\theta_{\rm VH}\to 0} \Delta R_{\rm NL}(x) = 
\frac{W}{2 \ell_{\rm v}}\sigma^2_{{\rm v}, xy} \rho^3_{{\rm c}, xx}e^{-|x|/\ell_{\rm v}}~.
\end{equation}
This coincides with the available results in the literature~\cite{gorbachev_science_2014,abanin_prb_2009}. For a fixed value of the valley diffusion length~\cite{gorbachev_science_2014,sui_naturephys_2015,shimazaki_naturephys_2015}, Eq.~(\ref{eq:small_theta}) predicts  a cubic power-law scaling $\Delta R_{\rm NL}(x)\propto \rho^3_{{\rm c}, xx}$.

Upon increasing $\theta_{\rm VH}$, however, 
the dependence of $R_{\rm NL}(x)$ on $\rho_{{\rm c}, xx}$ significantly weakens. Asymptotically, for $\theta_{\rm VH} \to \pi/2$ (i.e.~for $\sigma_{{\rm v}, xy}\gg \sigma_{{\rm c}, xx}$), the nonlocal resistance (\ref{eq:non-local-valley-signal-small-Wtilde}) reduces to
\begin{equation}\label{eq:large_theta}
\lim_{\theta_{\rm VH}\to \pi/2} \Delta R_{\rm NL}(x) =\frac{W}{2 \ell_{\rm v}}\frac{1}{\sigma_{{\rm v}, xy}}~.
\end{equation}
Eq.~(\ref{eq:large_theta}) implies that, for large VHAs, the nonlocal resistance $\Delta R_{\rm NL}(x)$ due to bulk topological valley transport becomes {\it independent} of $\rho_{{\rm c}, xx}$, provided that one takes $\ell_{\rm v}$ to be constant~\cite{gorbachev_science_2014,abanin_prb_2009}. At the same time, 
$\Delta R_{\rm NL}(x)$ spreads out in space to very large distances, since the renormalized valley diffusion length (\ref{eq:effective-diffusion-length}) formally diverges.

In summary, we have presented a solution of the nonlocal topological valley transport problem that transcends the usual approximation~\cite{abanin_prb_2009} $\sigma_{{\rm v}, xy} \ll \sigma_{{\rm c}, xx}$. We have obtained an expression for the nonlocal resistance $R_{\rm NL}(x)$---Eq.~(\ref{eq:nonlocal-resistance})---which is valid for arbitrary values of the valley Hall angle $\tan{(\theta_{\rm VH})} = \sigma_{{\rm v}, xy}/\sigma_{{\rm c}, xx}$. For distances $x$ that are much larger than the valley diffusion length $\ell_{\rm v}$ and for $\ell_{\rm v}$ much larger than the van der Pauw scale $W/\pi$, we have obtained a compact formula---Eq.~(\ref{eq:non-local-valley-signal-small-Wtilde})---for the nonlocal resistance, which is again valid for arbitrary values of the valley Hall angle. 

The saturation observed in Refs.~\onlinecite{sui_naturephys_2015,shimazaki_naturephys_2015}, where the gap was increased by increasing the applied external electric field perpendicular to the van der Waals stack, is compatible with our findings.  Indeed~\cite{shimazaki_naturephys_2015}, for $W=1~{\rm \mu m}$ and 
$\sigma_{{\rm v},xy} = 4e^2/h$, Eq.~(\ref{eq:large_theta}) yields $\Delta R_{\rm NL} \simeq 500~{\rm \Omega}$ for $\ell_{\rm v} = 6.5~{\rm \mu m}$.

\acknowledgments
This work was supported by Fondazione Istituto Italiano di Tecnologia and the European Union's Horizon 2020 research and innovation programme under grant agreement No.~696656 ``GrapheneCore1''. It is a great pleasure to acknowledge A.K. Geim for inspiring discussions.

\end{document}